\documentclass[preprint,review,11pt]{elsarticle}
\pdfoutput=1
\usepackage{amssymb}
\usepackage{amsthm}
\usepackage{epsfig}
\usepackage[running,mathlines]{lineno}
\usepackage{amsmath}
\usepackage[colorlinks]{hyperref}
\usepackage{amssymb}
\usepackage{lipsum}
\usepackage{footnote}
\usepackage{soul,xcolor}
\usepackage{graphicx}
\usepackage{array, makecell} 
\usepackage{bm}
\usepackage{gensymb}
\usepackage{enumitem}
\usepackage{float}

\DeclareMathOperator{\erf}{erf}
\journal{NIM A}

\begin{document}

\title{Measurements of electron transport in liquid and gas Xenon using a laser-driven photocathode}

\author[a]{O.~Njoya,}
\ead{oumarou.njoya@stonybrook.edu}
\author[b]{T.~Tsang,}
\author[a, 1]{M.~Tarka,}
\fntext[1]{now at Physics Department, University of Massachusetts, Amherst, MA, USA}
\author[c]{W.~Fairbank,}
\author[a, 1]{K.S.~Kumar,}
\author[b]{T.~Rao}
\author[c]{T.~Wager}
\author[e]{S.~Al Kharusi,}
\author[f]{G.~Anton,}
\author[g]{I.J.~Arnquist,}
\author[h, 2]{I.~Badhrees,}
\fntext[2]{also at King Abdulaziz City for Science and Technology, Riyadh, Saudi Arabia}
\author[i]{P.S.~Barbeau,}
\author[j]{D.~Beck,}
\author[k]{V.~Belov,}
\author[l]{T.~Bhatta,}
\author[m]{J.P.~Brodsky,}
\author[n]{E.~Brown,}
\author[e,o]{T.~Brunner,}
\author[q, 3]{E.~Caden,}
\fntext[3]{also at SNOLAB, Sudbury, Ontario, Canada}
\author[s, 4]{G.F.~Cao,}
\fntext[4]{also at University of Chinese Academy of Sciences, Beijing, China}
\author[p]{L.~Cao,}
\author[s]{W.R.~Cen,}
\author[c, 5]{C.~Chambers,}
\fntext[5]{now at Physics Department, McGill University, Montr\'eal, Qu\'ebec, Canada}
\author[h]{B.~Chana,}
\author[r]{S.A.~Charlebois,}
\author[b]{M.~Chiu,}
\author[q, 3]{B.~Cleveland,}
\author[j]{M.~Coon,}
\author[c]{A.~Craycraft,}
\author[t]{J.~Dalmasson,}
\author[u]{T.~Daniels,}
\author[e]{L.~Darroch,}
\author[v]{S.J.~Daugherty,}
\author[w,o]{A.~De St. Croix,}
\author[q]{A.~Der Mesrobian-Kabakian,}
\author[t]{R.~DeVoe,}
\author[g]{M.L.~Di~Vacri,}
\author[o,w]{J.~Dilling,}
\author[s]{Y.Y.~Ding,}
\author[x]{M.J.~Dolinski,}
\author[d]{A.~Dragone,}
\author[j]{J.~Echevers,}
\author[h]{M.~Elbeltagi,}
\author[y]{L.~Fabris,}
\author[c]{D.~Fairbank,}
\author[q]{J.~Farine,}
\author[g]{S.~Ferrara,}
\author[z]{S.~Feyzbakhsh,}
\author[r]{R.~Fontaine,}
\author[n]{A.~Fucarino,}
\author[w,o]{G.~Gallina,}
\author[x]{P.~Gautam,}
\author[b]{G.~Giacomini,}
\author[h]{D.~Goeldi,}
\author[h,o]{R.~Gornea,}
\author[t]{G.~Gratta,}
\author[x]{E.V.~Hansen,}
\author[m]{M.~Heffner,}
\author[g]{E.W.~Hoppe,}
\author[f]{J.~H\"{o}{\ss}l,}
\author[m]{A.~House,}
\author[aa]{M.~Hughes,}
\author[c]{A.~Iverson,}
\author[bb]{A.~Jamil,}
\author[t]{M.J.~Jewell,}
\author[s]{X.S.~Jiang,}
\author[k]{A.~Karelin,}
\author[d, 6]{L.J.~Kaufman,}
\fntext[6]{also at Indiana University, Bloomington, IN, USA}
\author[z]{D.~Kodroff,}
\author[h]{T.~Koffas,}
\author[w,o]{R.~Kr\"{u}cken,}
\author[k]{A.~Kuchenkov,}
\author[w,o]{Y.~Lan,}
\author[l]{A.~Larson,}
\author[cc]{K.G.~Leach,}
\author[t]{B.G.~Lenardo,}
\author[dd]{D.S.~Leonard,}
\author[t]{G.~Li,}
\author[j]{S.~Li,}
\author[bb]{Z.~Li,}
\author[q]{C.~Licciardi,}
\author[x]{Y.H.~Lin,}
\author[s]{P.~Lv,}
\author[l]{R.~MacLellan,}
\author[e]{T.~McElroy,}
\author[e]{M.~Medina-Peregrina,}
\author[f]{T.~Michel,}
\author[d]{B.~Mong,}
\author[bb]{D.C.~Moore,}
\author[e]{K.~Murray,}
\author[aa]{P.~Nakarmi,}
\author[cc]{C.R.~Natzke,}
\author[y]{R.J.~Newby,}
\author[b]{Z.~Ning,}
\author[r]{F.~Nolet,}
\author[aa]{O.~Nusair,}
\author[n]{K.~Odgers,}
\author[d]{A.~Odian,}
\author[d]{M.~Oriunno,}
\author[g]{J.L.~Orrell,}
\author[g]{G.S.~Ortega,}
\author[aa]{I.~Ostrovskiy,}
\author[g]{C.T.~Overman,}
\author[r]{S.~Parent,}
\author[aa]{A.~Piepke,}
\author[z]{A.~Pocar,}
\author[r]{J.-F.~Pratte,}
\author[b]{V.~Radeka,}
\author[b]{E.~Raguzin,}
\author[b]{S.~Rescia,}
\author[o]{F.~Reti\`{e}re,}
\author[x]{M.~Richman,}
\author[q]{A.~Robinson,}
\author[r]{T.~Rossignol,}
\author[d]{P.C.~Rowson,}
\author[r]{N.~Roy,}
\author[i]{J.~Runge,}
\author[g]{R.~Saldanha,}
\author[m]{S.~Sangiorgio,}
\author[d]{K.~Skarpaas~VIII,}
\author[aa]{A.K.~Soma,}
\author[r]{G.~St-Hilaire,}
\author[k]{V.~Stekhanov,}
\author[m]{T.~Stiegler,}
\author[s]{X.L.~Sun,}
\author[c]{J.~Todd,}
\author[s, 7]{T.~Tolba,}
\fntext[7]{now at at IKP, Forschungszentrum J\"ulich, Germany}
\author[e]{T.I.~Totev,}
\author[g]{R.~Tsang,}
\author[r]{F.~Vachon,}
\author[aa]{V.~Veeraraghavan,}
\author[h]{S.~Viel,}
\author[v]{G.~Visser,}
\author[h]{C.~Vivo-Vilches,}
\author[ee]{J.-L.~Vuilleumier,}
\author[f]{M.~Wagenpfeil,}
\author[q]{M.~Walent,}
\author[p]{Q.~Wang,}
\author[o, 8]{M.~Ward,}
\fntext[8]{now at Department of Physics, Queen’s University, Kingston, Ontario, Canada}
\author[h]{J.~Watkins,}
\author[t]{M.~Weber,}
\author[s]{W.~Wei,}
\author[s]{L.J.~Wen,}
\author[q]{U.~Wichoski,}
\author[t]{S.X.~Wu,}
\author[s]{W.H.~Wu,}
\author[p]{X.~Wu,}
\author[bb]{Q.~Xia,}
\author[p]{H.~Yang,}
\author[j]{L.~Yang,}
\author[x]{Y.-R.~Yen,}
\author[k]{O.~Zeldovich,}
\author[s]{J.~Zhao,}
\author[p]{Y.~Zhou}
\author[f]{and T.~Ziegler}

\address[a]{Department of Physics and Astronomy, Stony Brook University, SUNY, Stony Brook, NY 11794, USA}
\address[b]{Brookhaven National Laboratory, Upton, NY 11973, USA}
\address[c]{Physics Department, Colorado State University, Fort Collins, CO 80523, USA}
\address[d]{SLAC National Accelerator Laboratory, Menlo Park, CA 94025, USA}
\address[e]{Physics Department, McGill University, Montr\'eal, Qu\'ebec H3A 2T8, Canada}
\address[f]{Erlangen Centre for Astroparticle Physics (ECAP), Friedrich-Alexander University Erlangen-N\"urnberg, Erlangen 91058, Germany}
\address[g]{Pacific Northwest National Laboratory, Richland, WA 99352, USA}
\address[h]{Department of Physics, Carleton University, Ottawa, Ontario K1S 5B6, Canada}
\address[i]{Department of Physics, Duke University, and Triangle Universities Nuclear Laboratory (TUNL), Durham, NC 27708, USA}
\address[j]{Physics Department, University of Illinois, Urbana-Champaign, IL 61801, USA}
\address[k]{Institute for Theoretical and Experimental Physics named by A. I. Alikhanov of National Research Center ``Kurchatov Institute'', Moscow 117218, Russia}
\address[l]{Department of Physics, University of South Dakota, Vermillion, SD 57069, USA}
\address[m]{Lawrence Livermore National Laboratory, Livermore, CA 94550, USA}
\address[n]{Department of Physics, Applied Physics and Astronomy, Rensselaer Polytechnic Institute, Troy, NY 12180, USA}
\address[o]{TRIUMF, Vancouver, British Columbia V6T 2A3, Canada}
\address[p]{Institute of Microelectronics, Chinese Academy of Sciences, Beijing 100029, China}
\address[q]{Department of Physics, Laurentian University, Sudbury, Ontario P3E 2C6 Canada}
\address[r]{Universit\'e de Sherbrooke, Sherbrooke, Qu\'ebec J1K 2R1, Canada}
\address[s]{Institute of High Energy Physics, Chinese Academy of Sciences, Beijing 100049, China}
\address[t]{Physics Department, Stanford University, Stanford, CA 94305, USA}
\address[u]{Department of Physics and Physical Oceanography, University of North Carolina at Wilmington, Wilmington, NC 28403, USA}
\address[v]{Department of Physics and CEEM, Indiana University, Bloomington, IN 47405, USA}
\address[w]{Department of Physics and Astronomy, University of British Columbia, Vancouver, British Columbia V6T 1Z1, Canada}
\address[x]{Department of Physics, Drexel University, Philadelphia, PA 19104, USA}
\address[y]{Oak Ridge National Laboratory, Oak Ridge, TN 37831, USA}
\address[z]{Amherst Center for Fundamental Interactions and Physics Department, University of Massachusetts, Amherst, MA 01003, USA}
\address[aa]{Department of Physics and Astronomy, University of Alabama, Tuscaloosa, AL 35487, USA}
\address[bb]{Wright Laboratory, Department of Physics, Yale University, New Haven, CT 06511, USA}
\address[cc]{Department of Physics, Colorado School of Mines, Golden, CO 80401, USA}
\address[dd]{IBS Center for Underground Physics, Daejeon 34126, Korea}
\address[ee]{LHEP, Albert Einstein Center, University of Bern, Bern CH-3012, Switzerland}


\begin{abstract}
Measurements of electron drift properties in liquid and gaseous xenon are reported. The electrons are generated by the photoelectric effect in a semi-transparent gold photocathode driven in transmission mode with a pulsed ultraviolet laser. The charges drift and diffuse in a small chamber at various electric fields and a fixed drift distance of 2.0 cm. At an electric field of 0.5 kV/cm, the measured drift velocities and corresponding temperature coefficients respectively are $1.97 \pm 0.04$ mm/$\mu$s and $(-0.69\pm0.05)$\%/K for liquid xenon, and $1.42 \pm 0.03$ mm/$\mu$s and $(+0.11\pm0.01)$\%/K for gaseous xenon at 1.5 bar. In addition, we measure longitudinal diffusion coefficients of $25.7 \pm 4.6$ cm$^2$/s and $149 \pm 23$ cm$^2$/s, for liquid and gas, respectively. The quantum efficiency of the gold photocathode is studied at the photon energy of 4.73 eV in liquid and gaseous xenon, and vacuum. These charge transport properties and the behavior of photocathodes in a xenon environment are important in designing and calibrating future large scale noble liquid detectors.
\end{abstract}
\maketitle




\section{\label{sec:level1}Introduction}

In recent years liquid xenon (LXe) time projection chambers (TPCs) have proven to be excellent detectors in the searches for neutrinoless double beta decay \cite{albert2014search,albert2017searches} and dark matter \cite{aprile2016xenon100,aprile2016low,akerib2017results,aprile2017xenon1t} as well as for other low-background physics searches \cite{albert2017searches,albert2018searchnucleon}. Xenon is attractive because it can be chemically and radiologically purified to very high levels and its high density and atomic number provide substantial shielding against background radiation \cite{albert2014improved}. Detectors ranging from a few to hundreds of kilograms have produced high-quality results, paving the way for future tonne-scale detectors; XENONnT, LZ, and PandaX are multi-tonne LXe detectors for the direct detection of dark matter \cite{aprile2017xenon1t,collaboration2019observation,akerib2018projected,mount2017lux,cui2017dark}. nEXO, the proposed successor to EXO-200, is a ton-scale experiment that aims to perform a search for neutrinoless double beta decay of $^{136}$Xe, with a design half-life sensitivity of $\sim 10^{28}$ years \cite{albert2017sensitivity,kharusi2018nexo}. Proposals for large GXe TPCs for neutrinoless double beta decay of $^{136}$Xe have also been put forth.

A distinctive attribute of liquid noble elements, and xenon in particular, is the simultaneous production of ionization electrons and scintillation photons when exposed to ionizing radiation \cite{conti2003correlated,neilson2009characterization}. The longitudinal position of events in a LXe TPC is reconstructed using the delay between primary scintillation and the detection of ionization charge. However this can be complicated not only by electron losses but also by effects of electron diffusion which smear the spatial resolution of event localization; for this reason understanding electron diffusion is important. This is especially relevant when the drift distances are large (\textgreater 1 m) as in ton-scale experiments \cite{albert2017sensitivity,kharusi2018nexo,aprile2017xenon1t,mount2017lux}. Because the diffusion of electrons in high electric fields is generally anisotropic \cite{robson1972thermodynamic}, longitudinal (in the direction of the drift field) and transverse diffusion need to be measured separately. For most TPCs the longitudinal diffusion is smaller than the transverse diffusion due to the longitudinal confinement along the electric field \cite{aprile2010liquid}. Literature on measurement of the longitudinal diffusion of electric charges in LXe is sparse.

The inherent self-shielding of next generation LXe detectors presents new challenges for calibration and the monitoring of small time variations in LXe properties such as the electron lifetime. With a drift distance of 1.3 m in nEXO, an electron lifetime better than 10 ms is desired and the uncertainty in the lifetime correction must be at most 3\% \cite{albert2014improved,kharusi2018nexo}. \textit{In-situ} continuous monitoring of LXe properties will be one of the important factors to obtain optimal performance. The investigation of the feasibility of producing calibrated amounts of charge with laser pulses transported to gold photocathodes embedded in the main TPC cathode is an important facet of the nEXO R\&D effort. 

A gold cathode was used in a laboratory-scale setup as a laser-driven electron source to perform measurements of the longitudinal diffusion coefficient and drift speed, and their temperature dependence, of electrons as they drift in electric fields ranging from 70 V/cm to 1000 V/cm. These measurements are reported in this paper. Section \ref{sec:level2} contains a description of the experimental apparatus. In section \ref{sec:level3}, the data acquisition and analysis are described. The results are shown and discussed in section \ref{sec:level4} before concluding remarks in section \ref{sec:level5}.

\section{\label{sec:level2}Experimental Apparatus}

Three methods exist in the literature for the measurement of longitudinal electron diffusion: un-gridded \cite{hunter1986electron,kusano2012electron}, gridded \cite{davies1989measurements}, and shuttered \cite{takatou2011drift} drift cells. In this work, the gridded cell arrangement was employed because of its simplicity and similarity to LXe TPCs.

\begin{figure}[h]
	\centering
	\includegraphics[scale=0.45]{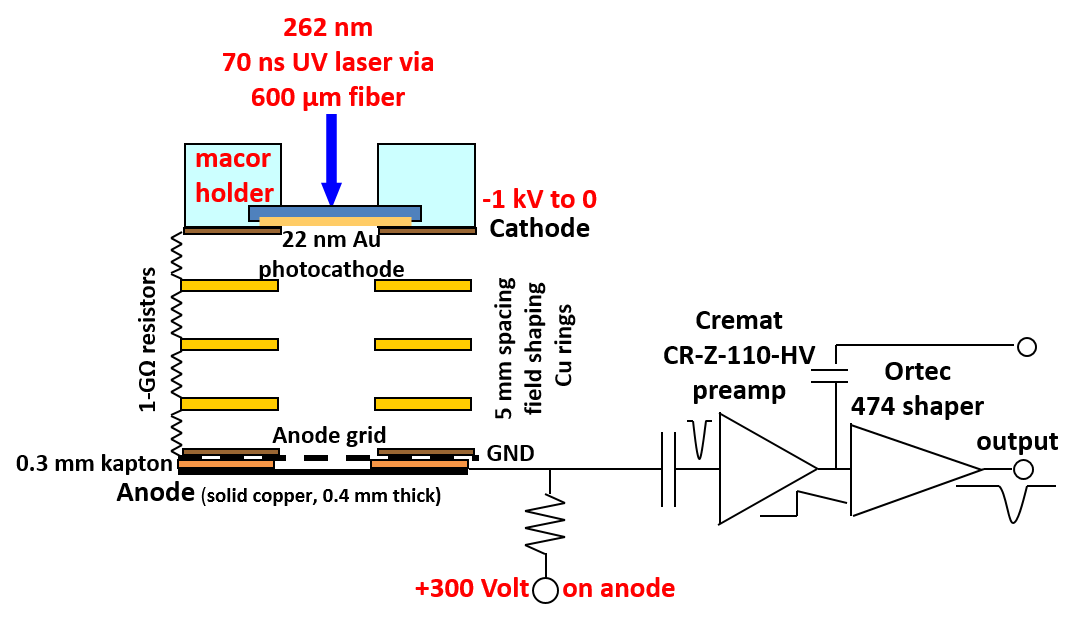}
	\caption{\label{fig:grid}A schematic diagram of the experimental setup. A UV laser back-illuminates a gold photocathode via a 600-$\mu$m fused-silica fiber. The space between the photocathode and the grid defines the drift region, where a uniform drift field is maintained with the help of copper field shaping rings. Electrons are collected on a Cu anode and their signal is amplified with the charge-sensitive preamplifier.}
\end{figure}

In this experiment, pulses of electrons are generated by back-illuminating a semitransparent gold photocathode with a pulsed UV laser. A schematic diagram of the drift chamber and accompanying electronics is shown in Figure \ref{fig:grid}. The drift chamber is housed in a 0.5 L cylinder with an inner diameter of 5 cm and a height of 27 cm. Elements of the cell include an optical fiber for photon delivery, a drift stack composed of a gold (Au) photocathode, three copper field shaping rings, and an anode grid followed by a copper anode. The photocathode is a 22-nm thick gold film ($\sim$50\% transmission at 266 nm) thermally evaporated on a 1-mm thick, 10-mm diameter sapphire disk. The sapphire disk sits in a macor holder at the top of the drift stack. The photon source is a pulsed frequency-quadrupled {(Spectra Physics: Evolution X,)} Nd:YLF laser with 4.73-eV photon energy and 71-ns pulse-width FWHM at 100-Hz repetition rate. The pulsed UV photons are coupled into a 4.35-m long 600-$\mu$m core diameter solarization-resistant UV fused silica optical fiber (ThorLabs: UM22-600) with a numerical aperture (NA) of 0.22 \cite{thorlabs}. One end of the portion of fiber that is in vacuum directly contacts the back surface of the sapphire disk. The laser light back-illuminates the Au photocathode (which has a work function of 4.2 eV in vacuum \cite{li2016measurement}), releasing photoelectrons into the LXe, where they drift along the lines of the uniform electric field. The UV laser energy per pulse is measured using a J3-02 Molectron pyroelectric energy detector fitted with a fiber-coupled adapter. The maximum laser energy deposited on the surface of the photocathode per pulse is $\sim$0.64 $\mu$J; this corresponds to an energy density of approximately 0.2 mJ/cm$^2$, far below the measured 6 mJ/cm$^2$ damage threshold of the thin gold film. The overall fiber transmission in the UV is $\sim$30\%. Therefore, after taking into account the $\sim$50\% attenuation of the photocathode, the optical throughput of the fiber to the surface of the photocathode is $\sim$15\% at UV wavelengths. 

A 280-$\mu$m thick phosphor-bronze disk with a clear inner diameter (ID) of 8 mm is negatively biased and makes electrical contact with the photocathode; it is securely fastened to a macor holder with ceramic screws. To ensure the uniformity of the drift field, three copper field shaping rings (2~mm thick, 21.3-mm outer diameter, 15.7-mm ID) are spaced at a 5-mm pitch between the cathode and the anode. These rings are precisely locked in place by four slotted alumina rods. The drift distance can be modified by adding or removing field shaping rings as needed and by shifting the cathode and anode assembly. A series of 1-G$\Omega$  cryogenic- and vacuum-compatible resistors electrically connect the photocathode, the field shaping rings, and the grounded anode grid. A 300-$\mu$m thick Kapton spacer is sandwiched between the anode grid and the anode. The bias voltage of +300 V is applied on the anode disk for the collection of drifted electrons. The grid (35-$\mu$m Ni-Cu wire width with 350-$\mu$m pitch) is mounted onto a 250-$\mu$m thick phosphor-bronze disk with an ID of 8 mm, and all three components are secured by ceramic screws to a macor holder at the bottom of the drift stack. Due to the low thermal expansion coefficients of macor and alumina, the change in drift distance (and hence the drift field) with temperature is negligible. The anode grid is kept at ground. The collection field---the field between the anode and the anode grid---is kept constant at 10000 V/cm to maximize transmission of electrons. The drift field between the cathode and the anode grid is varied from 70 V/cm to 1000 V/cm.

To perform the quantum efficiency measurements described in section \ref{sec:level4}-D, a second grid (not shown in Figure \ref{fig:grid}) identical to the anode grid is added in front the cathode. This grid is mounted on an additional phosphor-bronze plate and is separated from the cathode by a 1-mm thick macor insulator; the net distance from the surface of the cathode to the grid is calculated to be $1.28 \pm 0.02$ mm, accounting for the combined thickness of the plate between the cathode and the macor insulator. The extraction field between the photocathode surface and the cathode/upper grid is tuned to always be half the drift field. This is done to ensure optimal transmission through the grid \cite{bunemann1949design}.

Charges arriving at the anode are converted to voltage by a BNL IO535 or a Cremat-CR-Z-110-HV charge-sensitive preamplifier followed by an Ortec 474 timing filter amplifier with 100-ns shaping time. The preamplifier is calibrated by studying its response when a known amount of charge from a function generator is injected into it. Preamplifier and amplified shaped signals are both recorded on a 5~GS/s Agilent digital oscilloscope. The sweep trigger is the output of a \textless 1-ns risetime fast photodiode that intercepts a small portion of the frequency doubled Nd:YLF green laser beam. At the laser energy of $\sim$0.64 $\mu$J the typical number of drifting electrons ranges from $8\times10^4$ to $4\times10^5$ per pulse at bias fields of 70 to 1000 V/cm.

We employ all stainless steel pipes and valves in the xenon gas purification and recovery system. Furthermore, macor and alumina were chosen for drift stack construction to minimize out-gassing. All HV feedthroughs and thermocouples are Kapton-insulated. With these considerations and proper UHV handling techniques, vacuum levels in the low $10^{-6}$ torr range are achieved before the cell is filled with xenon.

The xenon is liquefied by immersing the cell in a cold ethanol bath (155 K). Two thermocouples are mounted in the cell, one near the anode and the other near the cathode, to monitor the temperature of the liquid; they also serve as level sensors during the initial fill-up (when the drift region is completely filled with LXe both thermocouples will read the same temperature). Liquefaction is also confirmed visually through a glass viewport located at the top of the cell. Note that during the gas measurements the cell can only be pressurized to a maximum of 2 bar because of the glass viewport. Prior to liquefaction, 99.999\% pure GXe is fed through a dry ice cold trap to remove water vapor, followed by a SAES purifier (a heated Zr/Al alloy getter capable of achieving ppb impurity levels)\cite{saes}. The purity level as determined from an estimate of the electron lifetime from a double-gridded measurement ranges from a few $\mu$s to about 35 $\mu$s. At the end of each run the xenon is recovered via cryopumping to a clean stainless steel tank. There is no active gas xenon recirculation.

\section{\label{sec:level3}Raw Waveform Analysis and Systematic Uncertainties}

Waveforms of collected electron bunches are recorded for various drift fields, laser energies, and temperatures using the digital oscilloscope. These waveforms are analyzed and experimental parameters such as electron signal amplitude, delay, and width are extracted.
\begin{figure}[!h]
	\centering
	\includegraphics[scale=0.225]{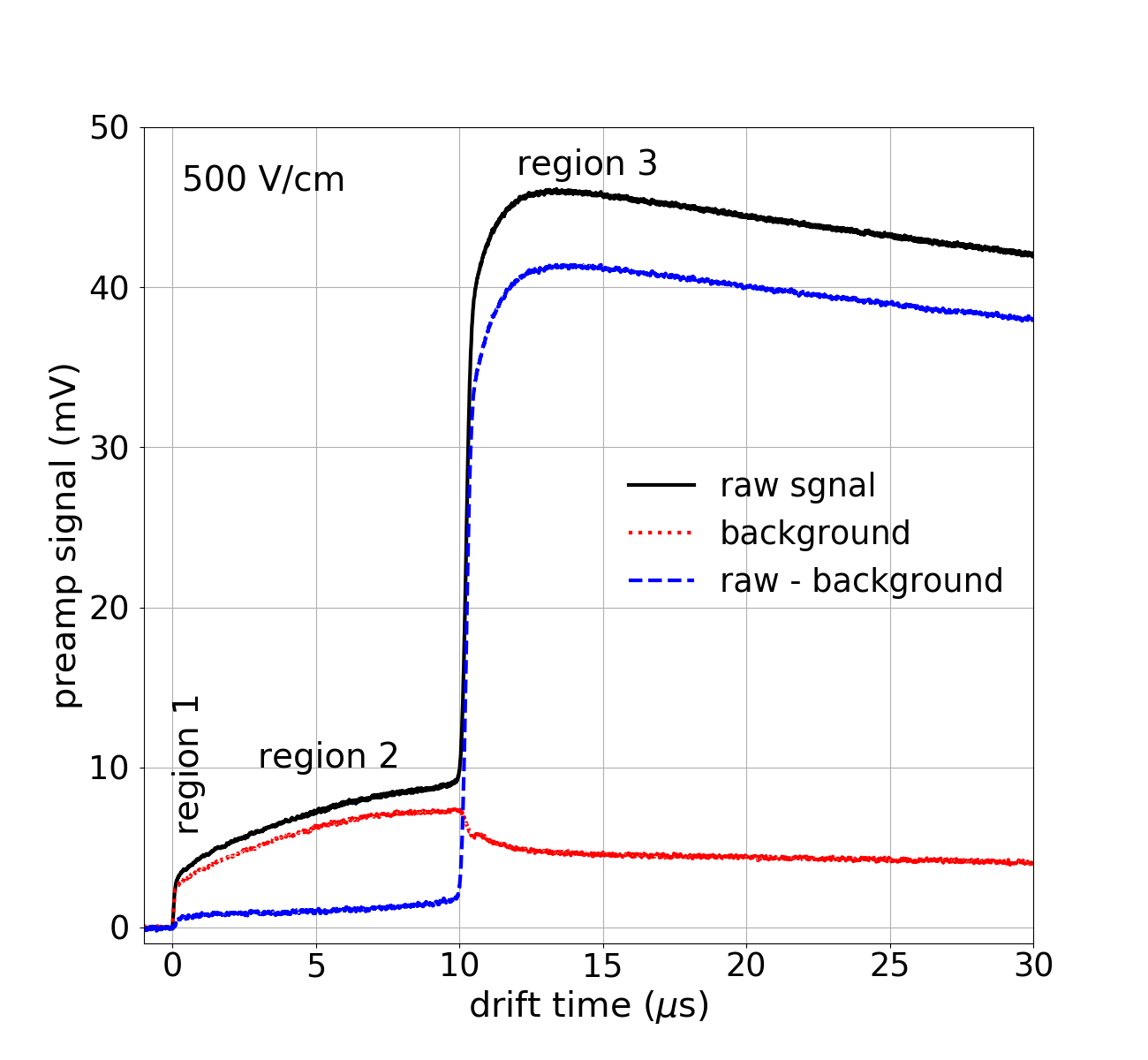}
	\caption{\label{fig:traces1}A typical raw preamplifier electron-charge signal trace (black), a background trace (red), and a background-subtracted trace (blue) in LXe are shown.}
\end{figure}

A typical raw preamplifier trace (black) is shown in Figure \ref{fig:traces1} along with a background (red) trace and a background-subtracted (blue) trace. The laser pulse trigger defines `time-zero' $t = 0$. The small rise of the signal near $t = 0$ within region 1 in Figure \ref{fig:traces1} is due to electrons generated at the anode grid by the $\sim$50\% laser light penetrating through the photocathode. This signal, which does not depend on the drift field and is proportional to the laser pulse energy, has no impact on the electrons generated from the photocathode.

The dominant step in the preamplifier signal occurs when the photoelectron bunch originating from the photocathode is collected by the anode. The time delay of the step and its risetime can be used to determine the drift time and the longitudinal width of the electron bunch, respectively. In the absence of longitudinal diffusion and with no Coulomb repulsion of the electron bunch, this step would have a fast risetime limited only by the response of the preamplifier convoluted with the initial electron bunch temporal width which is dictated by the laser pulse-width. The slow exponential decay (\textgreater 100 $\mu$s) is due to the RC time constant of the preamplifier.

The growth of the raw signal in region 2 of Figure \ref{fig:traces1} between 100 ns and the arrival time of the electrons at the anode indicates the presence of an additional background signal. There is qualitative evidence that this is the induction signal due to the motion of the electron bunch between the cathode and anode grid that results from imperfect shielding of the anode by the anode grid. To obtain a best approximation to the drift signal of the electron bunch region 3, the induction background is subtracted from the raw signal. The induction background signal is collected independently for each drift field by setting the anode bias voltage to 0V while maintaining the cathode at the same negative bias and the anode grid at ground; this voltage configuration ensures that a minimal amount of the electron bunch originating from the photocathode is collected at the anode.

\begin{figure}[!h]
	\centering
	\includegraphics[scale=0.225]{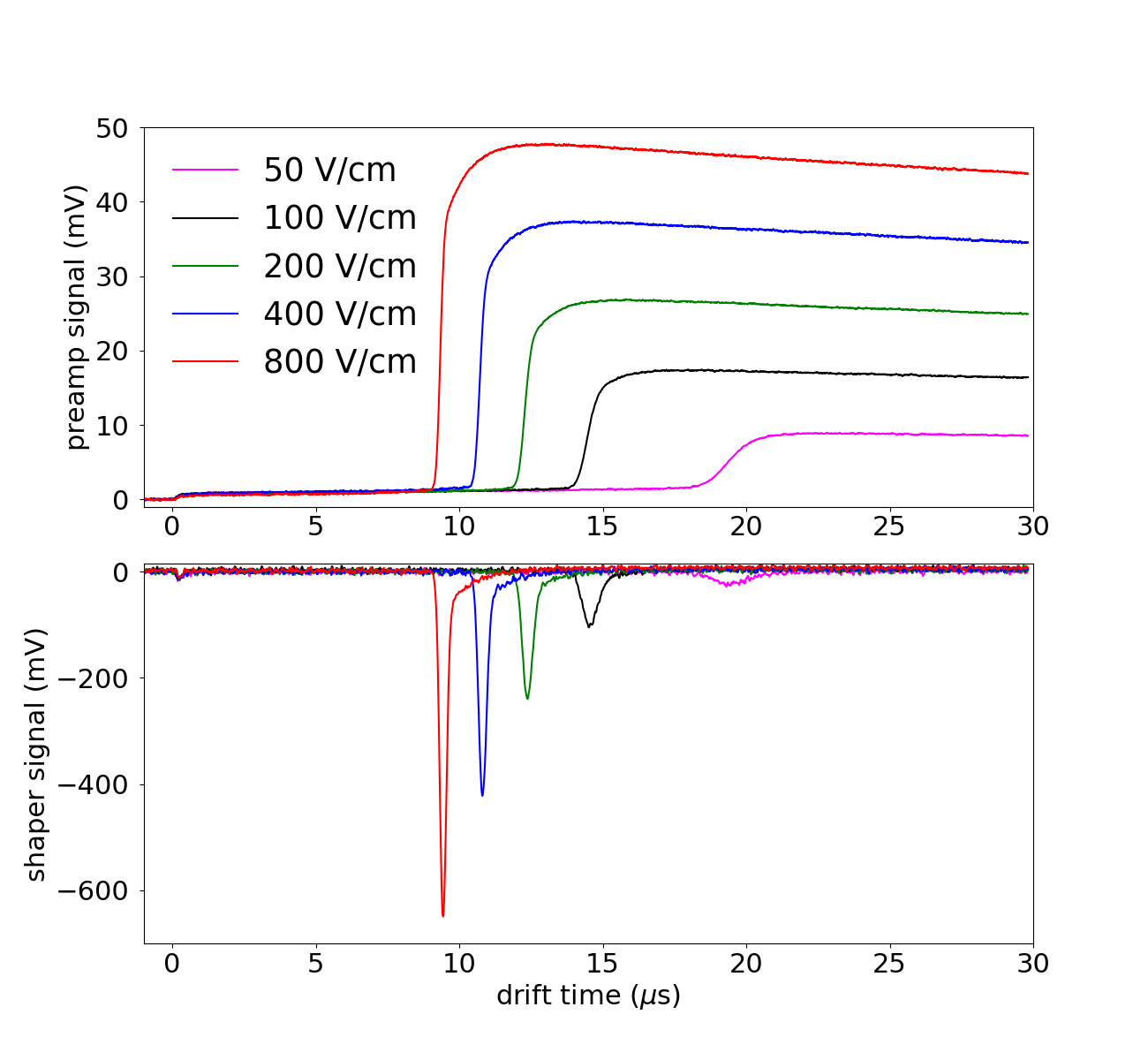}
	\caption{\label{fig:traces2}A representative set of background-subtracted preamplifier (top panel) and shaping amplifier (bottom panel) electron-charge signal traces at different drift fields in LXe is shown.}
\end{figure}

A set of background-subtracted preamplifier signal traces taken at various drift fields is shown in Figure \ref{fig:traces2}. It can be seen that decreasing drift fields are accompanied by increased drift times and longer risetimes. The reduction in signal height with lower drift field is the result of decreased quantum efficiency of the photocathode in LXe at lower electric fields. The longer risetime is a manifestation of increased electron bunch longitudinal spread at the anode that results primarily from longitudinal diffusion, as discussed in section \ref{sec:level4}-D.

Separate fits to the preamplifier and the shaping amplifier signals are performed for consistency. The preampflier trace is fitted with:

\begin{equation}
f(t,\sigma,\tau,A) = A \times (1+\erf(\frac{t-t_d}{\sigma\sqrt{2}})) \times \exp(-\frac{t-t_d}{\tau}),
\end{equation}
where $A$ is the amplitude, $t_d$ is the arrival time at the anode, the width $\sigma$ is the standard deviation of the Gaussian, and $\tau$ is the preamplifier RC time. The parameters $A$, $t_d$, and $\sigma$ for each waveform are determined by a least-squares fit. Similarly, the shaped signal is fitted with

\begin{equation}
g(t,\sigma,\tau,A) = A \times \exp(-\frac{(t-t_d)^{2}}{2\sigma^{2}}) \times \exp(-\frac{t-t_d}{\tau}).
\end{equation}
The preamplifier and the shaping-amplifier fit results agree to within 0.8\% for the time delays and 1.1\% for the widths.

The drift time is calculated by taking the difference between the arrival time at the anode $t_d$ and an offset of $t_0 = 47 \pm 7$ ns. This offset time accounts for the travel delay in the optical fiber as well as the combined instrument delay of the preamplifier, oscilloscope, and cable mismatch, and is derived from the photoemission signal in vacuum. There is an additional delay time from the anode grid to the anode of 0.1 $\mu$s which is treated as a systematic error. The value of 0.1 $\mu$s is obtained by assuming a drift speed of 3 mm/$\mu$s at 10000 V/cm \cite{yoshino1976effect}. The resulting error ranges from 0.4\% to 1.2\% depending on the drift field. This is the largest systematic error on the drift time.

Other sources of systematic errors were considered and are summarized in Table \ref{tab:table1}. In the case of the electron signal width $\sigma$ the largest average systematic error contribution (6.6\%) comes from the background subtraction. This value is the difference between $\sigma$ extracted from the raw traces and $\sigma$ from the background-subtracted traces; it thus accounts for any potential error in signal width measurements due to the background subtraction. 

Temperature variations between cathode and anode are maintained to less than 0.3 K and contribute less than 1\% uncertainty to the delay and the width of the signal. The impact of temperature is discussed in detail in section \ref{sec:level4}-C. 

The laser energy was found to have a significant impact on the signal width in LXe due to electron Coulomb repulsion and will be discussed in section \ref{sec:level4}-B. The width uncertainty associated with a correction applied for Coulomb repulsion is listed in Table \ref{tab:table1}.
\begin{table}[!h]
	\centering
	\caption{\label{tab:table1}%
		Summary of systematic uncertainties.
	}
	\setcellgapes{1.5pt}\makegapedcells
		\begin{tabular}{lcc}
			\textbf{error source}&
			\textbf{\makecell{Electron signal\\ delay ($t_d$)}}&
			\textbf{\makecell{Electron signal\\ width ($\sigma$)}}\\
			laser shot to shot fluctuations & 0.11\% & 2.8\% \\
			background subtraction & 0.1\% & 6.6\% \\
			anode grid to anode distance & 0.4-1.2\% & NA \\
			temperature & 0.36\%/K &  NA \\
			waveform model error & 0\% & 0.5\% \\
			Coulomb repulsion & $<$ 0.1\% & 1.0-4.1\%\\
		\end{tabular}
\end{table}

\section{\label{sec:level4}Results and Discussion}

\subsection{Electron Drift Velocity}
\begin{figure}[!h]
	\centering
	\includegraphics[scale=0.3]{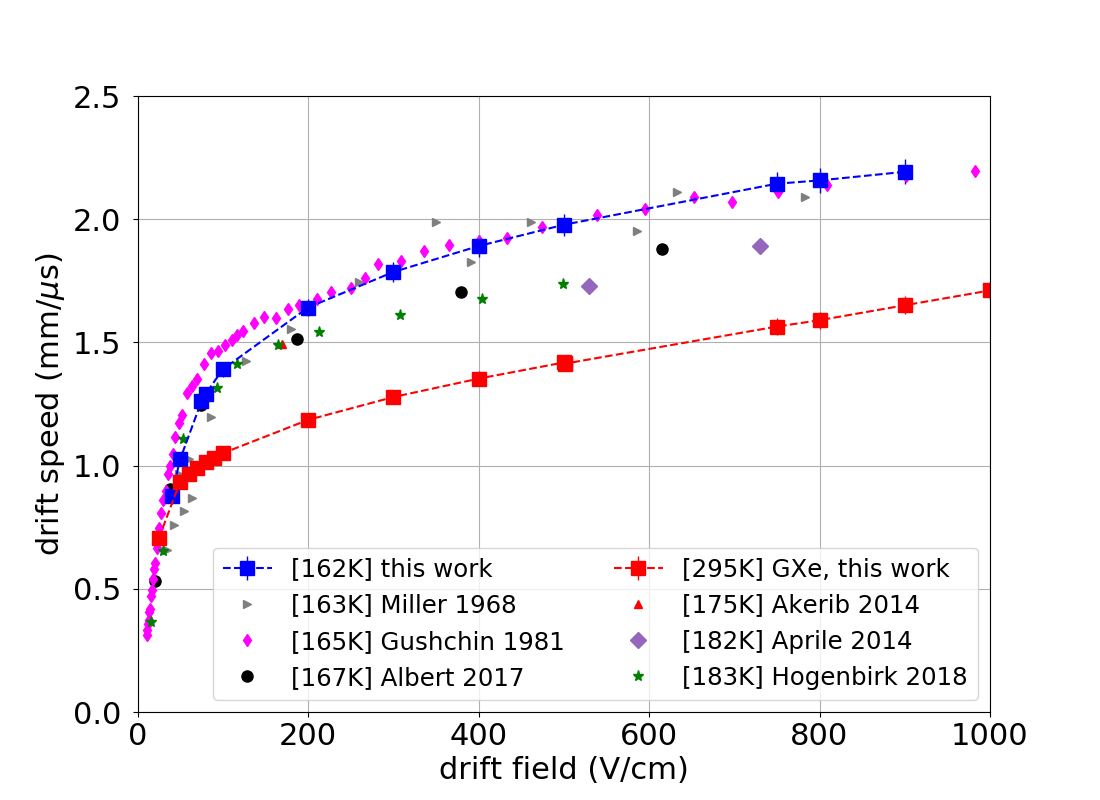}
	\caption{\label{fig:vvsE}The field dependence of electron drift velocity in LXe is shown. The present work is shown in blue squares with error bars. Other published values are also displayed with measurement temperature: EXO-200 (Albert \cite{albert2017measurement}), XENON100 (Aprile \cite{aprile2014analysis}), LUX (Akerib \cite{akerib2014first}), Gushchin et al. \cite{gushchin1982electron}, and Miller et al \cite{miller1968charge}. Measurements of drift velocity in GXe are also shown for comparison (red squares).}
\end{figure}

The drift velocity is given by

\begin{equation}
v = \frac{d}{t},
\end{equation}
where $d$ is the drift distance and $t=t_d-t_0$ is the drift time. Here the drift distance between the cathode and the anode grid mesh is $d = 20.0 \pm 0.1$ mm. Figure \ref{fig:vvsE} shows the measured electron drift velocity as a function of drift field in LXe and GXe. The error bars include statistical (average of multiple runs) and the systematic uncertainties described in the previous section and summarized in Table \ref{tab:table1}. At a drift field of 500 V/cm the measured drift velocity is 1.97 $\pm$ 0.04 mm/$\mu$s in LXe and 1.42 $\pm$ 0.03 mm/$\mu$s in GXe. It is evident that electrons drift faster in LXe than in GXe, which is in agreement with the literature \cite{aprile2010liquid,miller1968charge,yoshino1976effect,pack1992longitudinal}.

Previous reported measurements \cite{miller1968charge,gushchin1982electron,aprile2014analysis,akerib2014first,albert2017measurement} are also shown in Figure \ref{fig:vvsE}. The variance in reported electron drift velocity values for LXe at a given electric field is not well understood. The weak temperature dependence of the LXe drift velocity, discussed in section \ref{sec:level4}-C, is insufficient to account for the spread in reported measurements.

\begin{figure}[!h]
	\centering
	\includegraphics[scale=0.3]{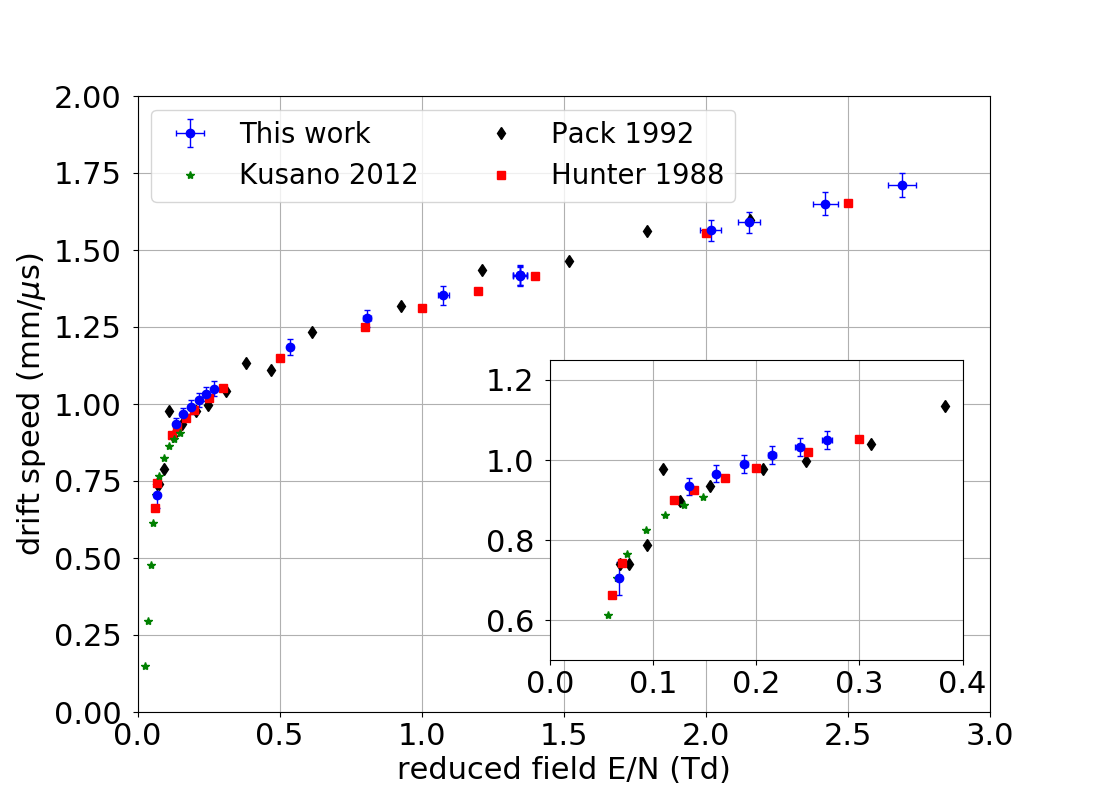}
	\caption{\label{fig:GvvsE}The electron drift velocity versus reduced field in GXe is shown. The reduced field, given by $E/N$ is in Townsend (Td) and $N$ is the GXe number density in the cell. Our measurements (blue circles) are in good agreement with those reported in \cite{kusano2012electron} (green crosses), \cite{pack1992longitudinal} (black diamonds), and \cite{hunter1988low} (red squares).}
\end{figure}

For GXe, the measured drift velocity plotted against the reduced field $E/N$ is shown in Figure \ref{fig:GvvsE} where $E$ is the drift field and $N$ is the GXe number density. The reduced field (in units of Townsend or 10$^{-17}$ Vcm$^{2}$) ranges from 0.1 Td to 2.6 Td. \cite{pack1992longitudinal}. The results reported here are in agreement with previously published measurements \cite{kusano2012electron,pack1992longitudinal,hunter1988low}.

\subsection{Longitudinal Diffusion}
A key feature of TPCs is the ability to accurately reconstruct events in 3D. The spread of the intrinsic electron bunch as it propagates due to diffusion can reduce the reconstruction accuracy thereby limiting the position resolution. The phenomenon is discussed in \cite{li2016measurement,agnes2018electroluminescence} and references therein. Transverse diffusion of electrons in LXe has been measured at a range of electric fields \cite{albert2017measurement,tadayoshi1982recent}. Four measurements of longitudinal diffusion in LXe have been reported in the literature \cite{sorensen2011anisotropic,mei2011thesis,shibumura2009,hogenbirk2018field}.

In the absence of Coulomb repulsion, the longitudinal diffusion coefficient $D_L$ in terms of the drift distance $d$, the drift time $t$, and the temporal width (standard deviation) in the longitudinal direction $\sigma_L$ is given by:

\begin{equation}
D_{L} = \frac{d^{2}\sigma_{L}^{2}}{2t^{3}}.
\label{eq:DL}
\end{equation}

In Equation \ref{eq:DL}, $\sigma_L$ is given by:

\begin{equation}
\sigma_L^2 = \sigma^2 - \sigma_0^2,
\label{eq:DL2}
\end{equation}
where $\sigma$ is the measured electron pulse width obtained by fitting the anode preamplifier signal and $\sigma_0$ is the initial broadening due to the laser pulse width (30 ns) and intrinsic preamplifier risetime of 35.3 $\pm$ 0.3 ns from fits to the calibrated preamplifier waveforms. This gives $\sigma_0 = 46.3 \pm 5.3$ ns.

Electrons are generated in bunches of about 10$^5$ or more before they begin to drift. Because of the short laser pulses used, the charge density is high enough in LXe that Coulomb repulsion becomes a significant factor in the growth of the bunch from its initial width to the width at the time of the measurement. To gauge this effect, the growth of the electron bunch is modeled as the electrons drift. This model is used to determine the Coulomb contribution to the measured $\sigma$.

Our simplified model of Coulomb interactions assumes an initial uniform charge distribution of ellipsoidal shape where the initial radius $w_{\parallel}$(0) along the longitudinal dimension is given by:

\begin{equation}
w_{\parallel}(t=0) = v(E) \times \Delta_{l},
\label{eq:w_para}
\end{equation}
where $v(E)$ is the electron drift speed at the applied drift field $E$ and $\Delta_l$ is the laser pulse temporal $1/e$ half-width. The initial transverse radius $w_{\perp}$(0) is the calculated laser spatial $1/e$ half-width after propagation through the 1-mm thick sapphire plate (on which the gold is evaporated). The propagation of such an ellipsoid in an electric field in vacuum is described in detail in \cite{grech2011coulomb}. The electrons at the front of the ellipsoid experience a greater longitudinal field $E+E_{\parallel}$, and move faster than the electrons at the back of the ellipsoid, which experience field $E-E_{\parallel}$ where $E_{\parallel}$ is the longitudinal Coulomb field at the surface of the ellipsoid. Thus the ellipsoid spreads under the combined effects of diffusion and Coulomb repulsion according to:

\begin{equation}
\frac{dw_{\parallel}}{dt} = {\beta}E_{\parallel}+\frac{2D_L}{w_{\parallel}},
\label{eq:CMw_para}
\end{equation}
\begin{equation}
\frac{dw_{\perp}}{dt} = {\mu}E_{\perp}+\frac{2D_T}{w_{\perp}},
\label{eq:CMw_perp}
\end{equation}
where $\beta = dv/dE$ is the slope of the drift velocity change at a particular E field, $E_{\perp}$ is the transverse Coulomb field, $\mu$ is the low-field electron mobility, and $D_T$ is the transverse diffusion coefficient. The mobility $\mu$ is calculated at low fields to approximate the fact that there is no drift field in the transverse direction. $\beta$ is calculated from the measured dependence of drift velocity on the E field (Figure \ref{fig:vvsE}). The diffusion terms in Equations \ref{eq:CMw_para} and \ref{eq:CMw_perp} account phenomenologically for radius change with time due to diffusion. $E_{\parallel}$ and $E_{\perp}$ are extracted from fits to the numerical simulations given in \cite{grech2011coulomb}:

\begin{equation}
E_{\perp} = \frac{3{\lambda}Q}{4{\pi\epsilon}w_{\perp}w_{\parallel}}\frac{0.5}{1+0.76\alpha}
\label{eq:CME_perp}
\end{equation}
and

\begin{equation}
E_{\parallel} = \frac{3{\lambda}Q}{4{\pi\epsilon}w_{\parallel}^2}(1-\frac{1}{1+1.54\alpha^{0.36}}),
\label{eq:CME_para}
\end{equation}
with $\alpha = w_{\perp}/w_{\parallel}$. $Q$ is the electron bunch total charge, $\lambda$ is a charge scale factor to account for differences between the uniform charge distribution with sharp edge of the model and the approximately Gaussian charge distribution of reality, and $\epsilon = \kappa\epsilon_0$ is the permittivity of LXe with dielectric constant $\kappa$ = 1.9 \cite{lide1995crc}. The temporal width $\sigma_{th}$ (standard deviation) of the ellipsoid at the anode is:

\begin{equation}
\sigma_{th} = \frac{w_{\parallel}(t)}{v\sqrt{2}}.
\end{equation}

The model outputs a $D_L$ value that minimizes the residuals between $\sigma_{th}$ (model) and $\sigma$ (data) at each drift field. The input parameters are $\beta$, $D_T$, $\mu$, $Q$, $\lambda$, and the drift time t. The charge $Q$ is set by the laser energy and applied drift field; $\beta$ and the drift time are set by the drift field. A range of $\mu$ values corresponding to previous drift speed measurements (see Figure \ref{fig:vvsE}) were tried and found to have no significant effect on $\sigma_{th}$. $D_T$ is taken from \cite{albert2017measurement} and has only a small effect on $\sigma_{th}$.

To determine the appropriate $\lambda$ value to use in the model, a SIMION \cite{simion} simulation of electron transport in GXe at 1.5 bar and 295K was compared to the Coulomb model for similar conditions. The simulation tracks the path of each electron with an average of 5 steps between collisions. The global Coulomb field of a specified number of electrons is included in the calculations. Gaussian-distributed electrons are released near the cathode and begin to drift. A low applied electric field of 50 V/cm was chosen so that a significant Coulomb repulsion effect could be seen in GXe. A collision cross section value of $1.2 \times 10^{-15}$ cm$^2$ was selected so that the drift time in the simulation agreed with the measured value at 500 V/cm. The transverse broadening in model and simulation were matched by adjusting the $D_T$ value in the Coulomb model.

\begin{figure}[h]
	\centering
	\includegraphics[scale=0.3]{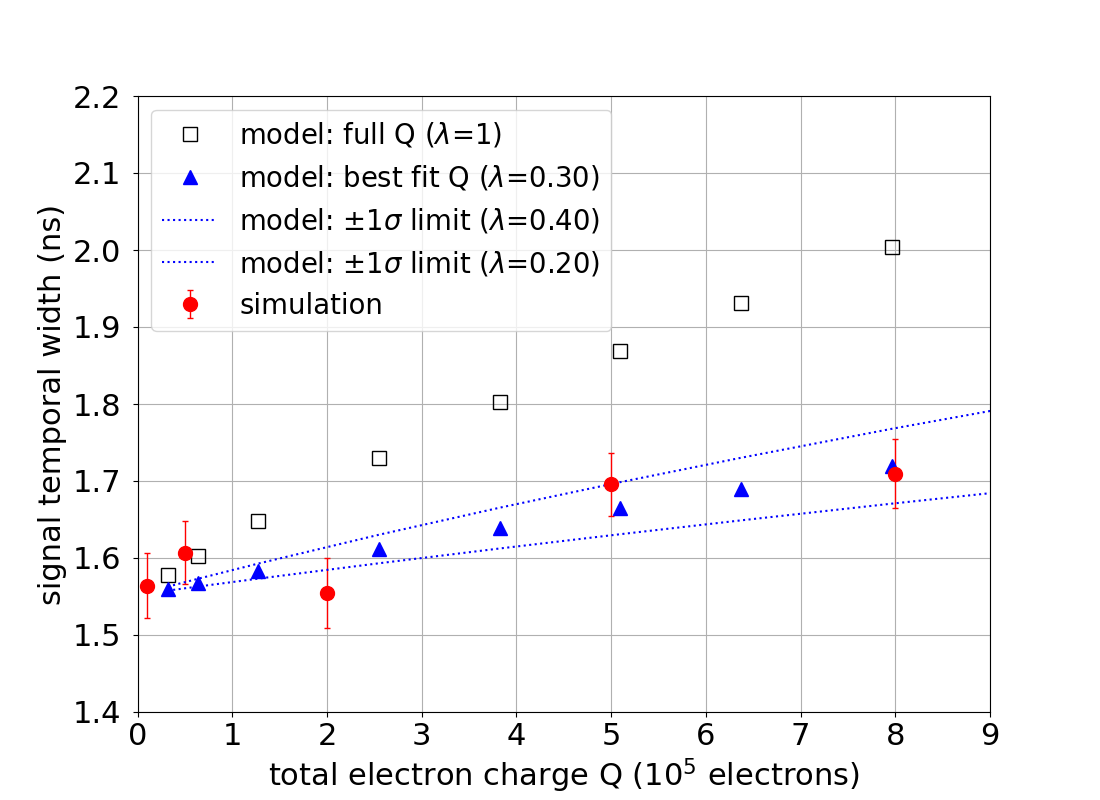}
	\caption{\label{fig:wvsLE2}Electron signal temporal widths versus total electron charge in GXe, for various charge fractions $\lambda$. The temporal widths calculated from the model (blue) include Coulomb repulsion and are in good agreement with those obtained from simulations (red).}
\end{figure}

For each of the four total charges simulated in GXe, the temporal width of the bunch and an error limit were determined by a Gaussian fit to a histogram of the electron arrival times at the anode. The results are shown in red circles in Figure \ref{fig:wvsLE2}. For Q $\sim$ 2$\times$10$^{5}$ electrons, the broadening due to Coulomb repulsion is not very large; however the broadening becomes more significant as Q increases to 8$\times$10$^{5}$ electrons. The model is fitted to the simulated GXe data by varying $D_L$ and $\lambda$ to minimize $\chi^2$. The best fit is shown by the solid blue triangles in Figure \ref{fig:wvsLE2}. The error limits on $\lambda$ are determined from the $\pm1\sigma$ limit on $\chi^2(\lambda)$. The model curves representing these error limits are shown by the $\pm1\sigma$ dashed blue lines in Figure \ref{fig:wvsLE2}. The Coulomb effect with the full charge $\lambda=1$ is clearly much larger than in the simulation. The result is $\lambda = 0.30\pm0.10$. For comparison, one might expect that the appropriate charge fraction to use in the simplified model, should correspond to that within the $1/e$ radius ($w$) of the Gaussian distribution. For an ellipsoidal Gaussian distribution, this charge fraction is 0.23.

The value $\lambda = 0.30 \pm 0.10$ is used to determine $D_L$ at each field in LXe by matching the model $\sigma_{th}$ to the experimental $\sigma$. The Coulomb correction (relative to Coulomb-free calculation of $D_L$) ranges from 6.1\% to 17.4\%. The Coulomb repulsion model error for $D_L$ at each drift field is determined from the error limits on $\lambda$ and ranges from 1.0\% to 4.1\% (last entry in Table \ref{tab:table1}).

\begin{figure}[!h]
	\centering
	\includegraphics[scale=0.3]{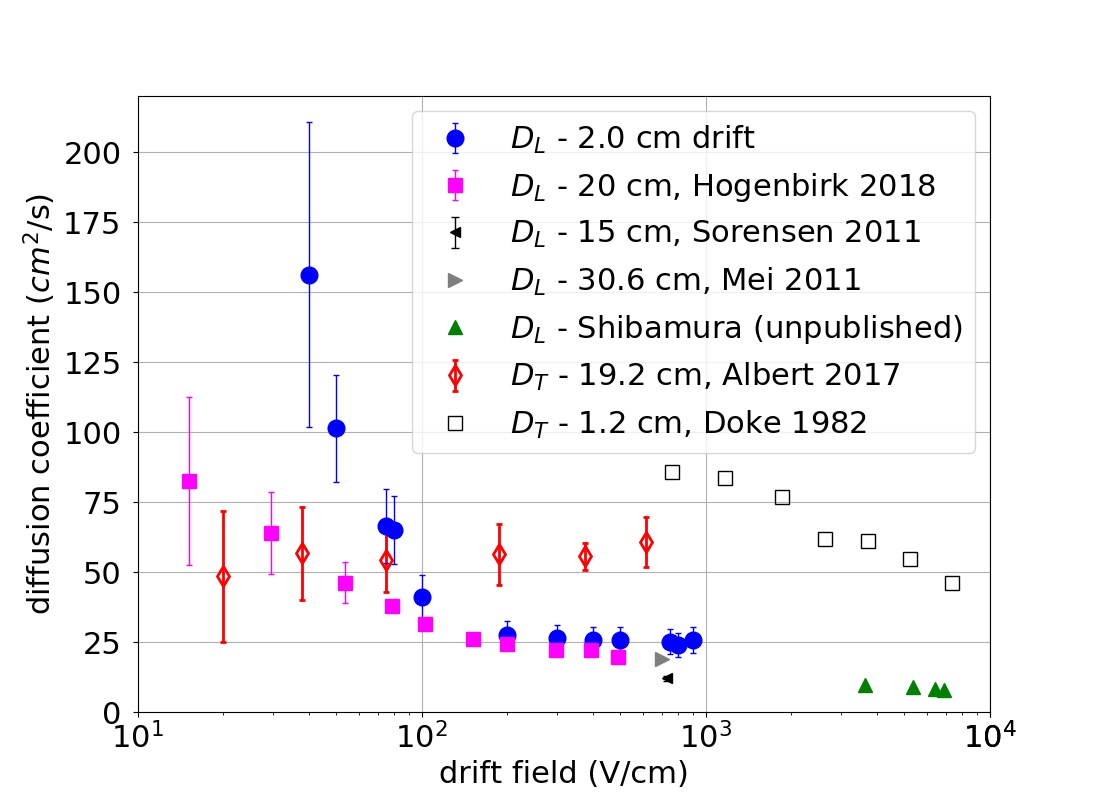}
	\caption{\label{fig:DLvsE} Electron longitudinal diffusion coefficient $D_L$ versus drift field in LXe. Values from this work (blue circles) and measurements from \cite{hogenbirk2018field} (magenta squares), \cite{sorensen2011anisotropic} (black triangle), \cite{mei2011thesis} (gray triangle), and Shibamura (green triangle) \cite{shibumura2009}. Also shown are the transverse diffusion coefficient $D_T$ from EXO-200 \cite{albert2017measurement} (hollow red diamonds), and \cite{tadayoshi1982recent} (hollow black squares).}
\end{figure}

The longitudinal diffusion coefficient $D_L$ in LXe is plotted as a function of drift field in Figure \ref{fig:DLvsE}. At 500 V/cm, $D_L = 25.7\pm4.6$ cm$^2$/s. The error bars include both statistical and systematic errors. Each data point on Figure \ref{fig:DLvsE} is an average of two distinct measurements. The statistical error is calculated as the standard deviation of the mean for the pair of measurements at each field.

Previous measurements of $D_L$ include a set of $D_L$ values for fields ranging from 15 V/cm to 493 V/cm \cite{hogenbirk2018field}, a single $D_L$ value at 700 V/cm \cite{mei2011thesis}, a single $D_L$ value at 730 V/cm \cite{sorensen2011anisotropic},  and a set of $D_L$ values for fields ranging from 3.6 kV/cm to 6.8 kV/cm \cite{shibumura2009}. These are shown in Figure \ref{fig:DLvsE}. Our measurements agree with those of \cite{hogenbirk2018field} at higher fields but are somewhat higher at low fields. For comparison, the transverse diffusion coefficient $D_T$ values at various fields are shown \cite{albert2017measurement,tadayoshi1982recent}. It is notable that $D_L$ rises as the field decreases and becomes comparable to $D_T$ at $\sim$ 100 V/cm.

\begin{figure}[!h]
	\centering
	\includegraphics[scale=0.3]{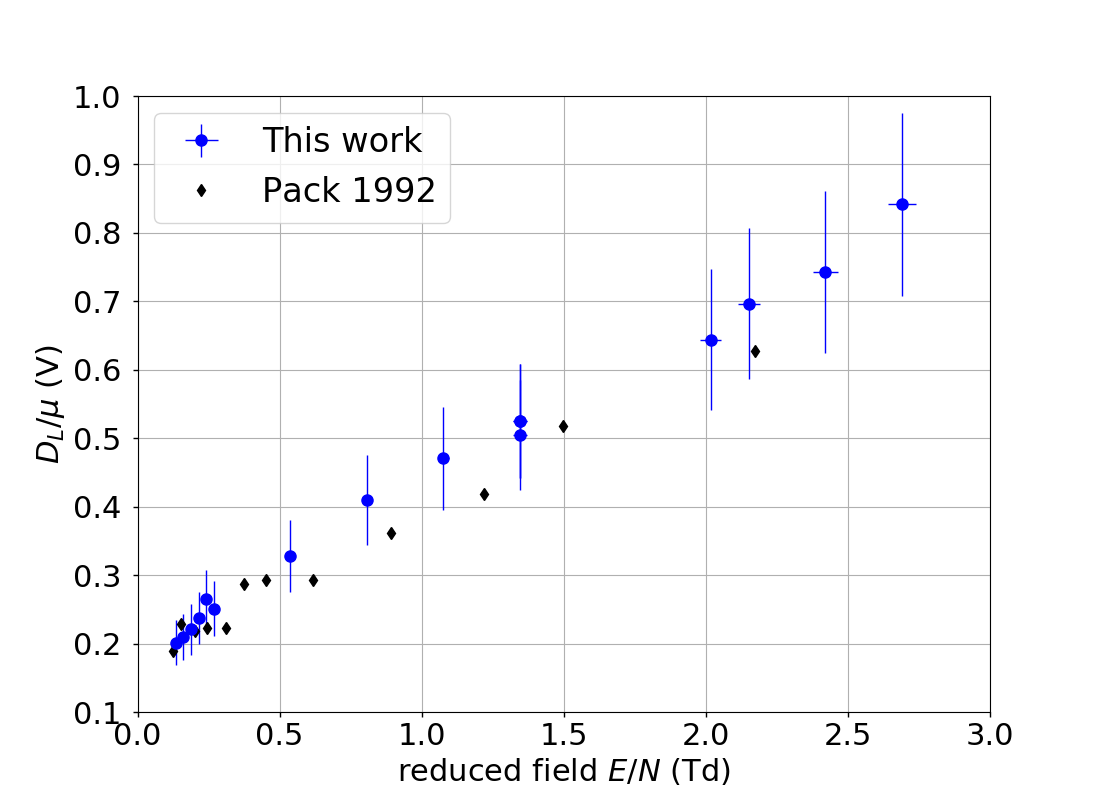}
	\caption{\label{fig:GDLvsE} GXe $D_L/\mu$ versus reduced field $E/N$ at room temperature is shown. The reduced field is given by $E/N$. Results from Pack \cite{pack1992longitudinal} are shown in black squares.}
\end{figure}

In the case of GXe, the measured ratio $D_L/\mu$ as a function of reduced field $E/N$, where $\mu = v/E$ is the electron mobility, is shown in Figure \ref{fig:GDLvsE}. Here $D_L$ = 149 $\pm$ 23 cm$^2$/s at 0.5 kV/cm. The Coulomb repulsion model gave no significant Coulomb correction for the GXe data. The values from \cite{pack1992longitudinal} are overlaid in Figure \ref{fig:GDLvsE} for comparison. The overall drift field dependence of $D_L/\mu$ for electrons in GXe are in good agreement with \cite{pack1992longitudinal}. Measurements at lower $E/N$ values have also been reported \cite{kusano2012electron,mcdonald2019electron}.

\subsection{Temperature Dependence}
\begin{figure}[!h]
	\centering
	\includegraphics[scale=0.20]{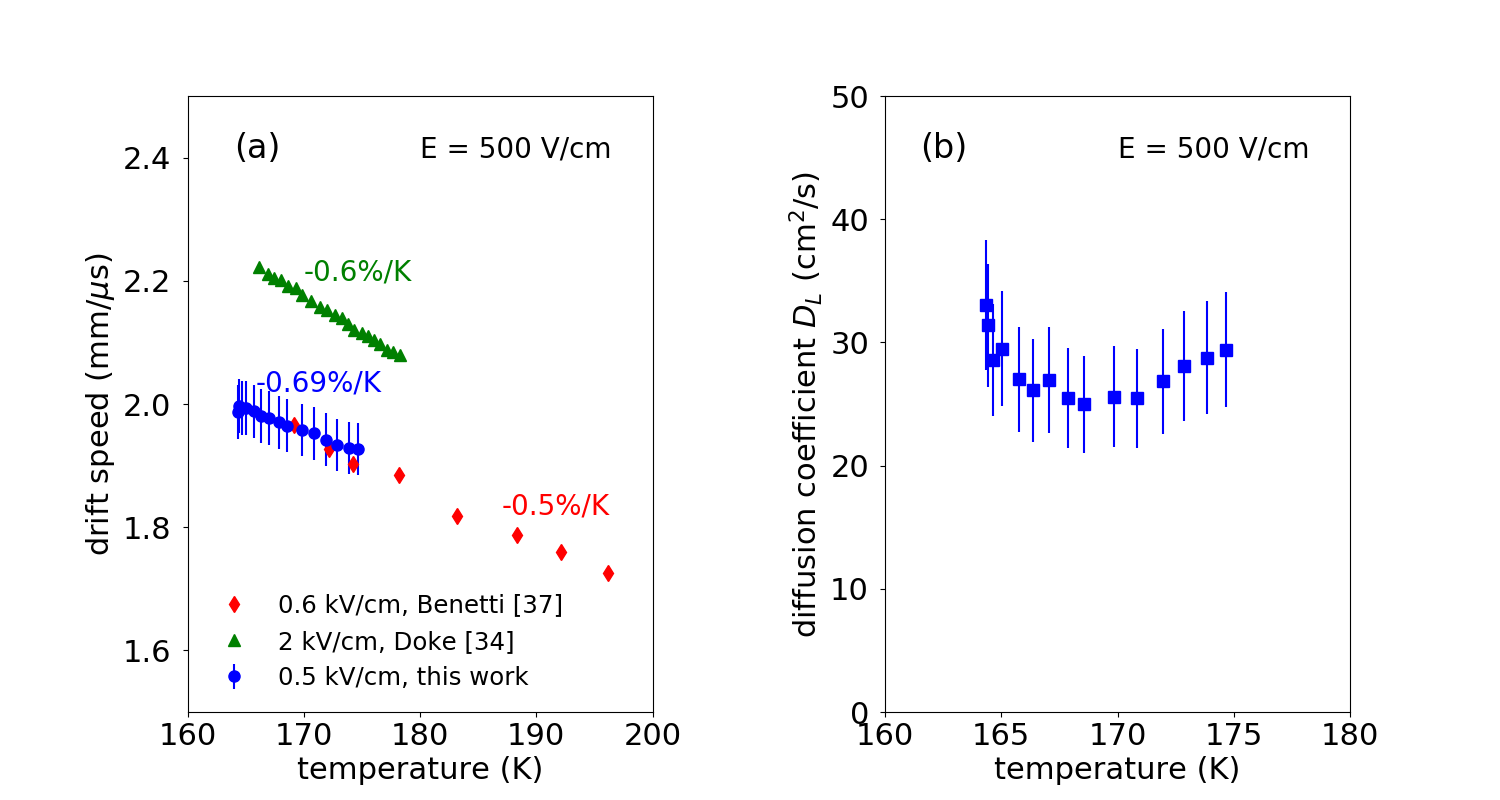}
	\caption{\label{fig:vvsT}(a) Temperature dependence of the electron drift velocity in LXe (blue circles). Also included are measurements of Doke (green pyramids) \cite{tadayoshi1982recent} and Benetti (red diamonds) \cite{benetti1993simple}. (b) Temperature dependence of the electron longitudinal diffusion coefficient in LXe (blue squares).}
\end{figure}

The temperature dependence of the drift speed and longitudinal diffusion coefficient were studied to help quantify the importance of the temperature uniformity inside a LXe TPC and assess whether it could explain the spread of drift velocity values found in the literature. These results are summarized in Figure \ref{fig:vvsT}. At the drift field of 500 V/cm, the electron drift speed in LXe decreases linearly with temperature at a rate of $-0.69 \pm 0.05$ \%/K, in qualitative agreement with previous measurements at various drift fields \cite{benetti1993simple,tadayoshi1982recent}. On the other hand, no significant temperature dependence of $D_L$ in LXe (also shown in Figure \ref{fig:vvsT}) is found within the uncertainties of the measurements.

\begin{figure}[!h]
	\centering
	\includegraphics[scale=0.3]{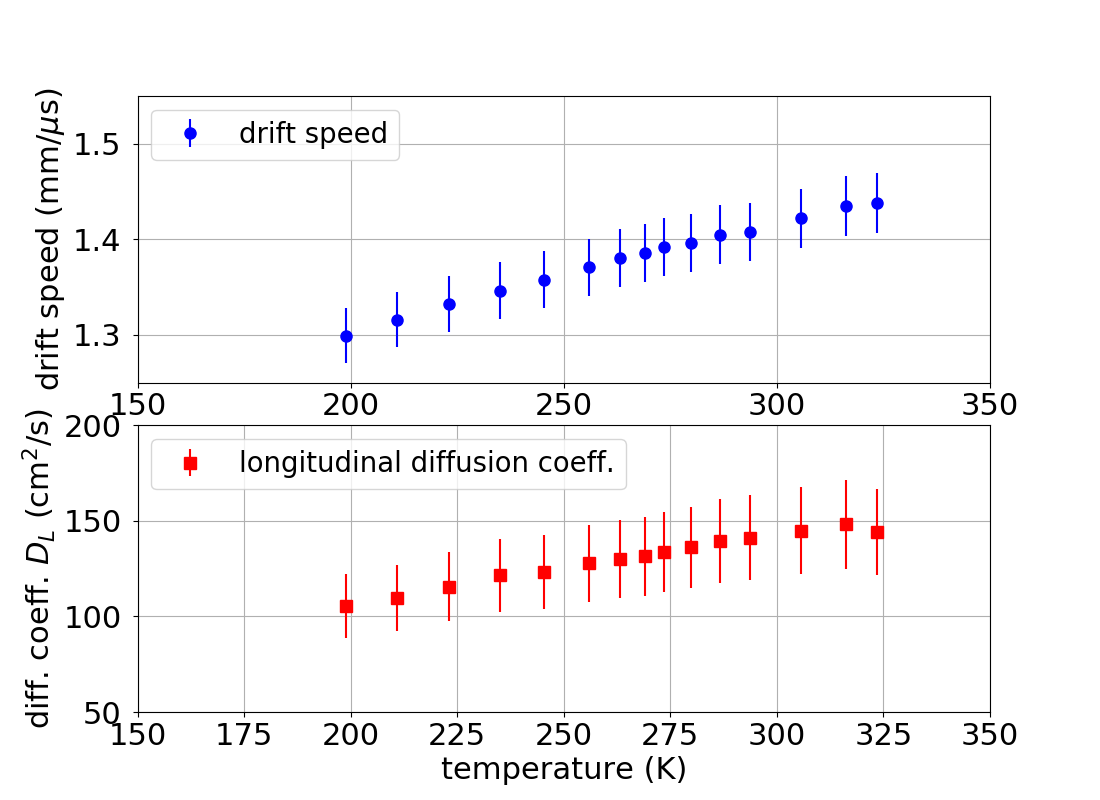}
	\caption{\label{fig:GvvsT}Temperature dependence of electron drift velocity (top panel) and longitudinal diffusion coefficient $D_L$ (bottom panel) in GXe at 500 V/cm drift field and pressure $P = 1.53 \pm 0.01$ bar.}
\end{figure}

In GXe, both $D_L$ and the drift velocity increase linearly with temperature; this is shown in Figure \ref{fig:GvvsT}. The drift velocity increases with temperature in GXe, while it decreases in LXe. A similar behavior has been reported in LAr and GAr \cite{li2016measurement}. 

\subsection{Quantum Efficiency of Au Photocathode}
The quantum efficiency ($QE$) is defined as the number of electrons leaving the photocathode per UV photon incident on the back surface of the photocathode. To perform this measurement, charges leaving the cathode were measured. This was enabled by the addition a mesh 1.28 $\pm$ 0.02 mm away from the cathode, as described in Section-\ref{sec:level2}. The $QE$ was studied as a function of extraction field in vacuum, room temperature GXe (1.5 bar), and in LXe. Due to changes in absolute photocathode response, possibly due to environmental changes, only qualitative behavior and ranges of values are reported. Measurements were made at extraction fields ranging from 25 V/cm to 1000 V/cm. In vacuum the measured $QE$ depends only very weakly on extraction field $E$; however it grows monotonically in both LXe and GXe.

QEs of $\sim 5 \times 10^{-6}$ are obtained in vacuum. This is consistent with values reported by \cite{li2016measurement} for gold photocathodes similar to the ones used here. The $QE$ in LXe and the $QE$ in GXe range between  $(1-5) \times 10^{-7}$. Remarkably, the $QE$ in LXe and GXe were nearly identical at each extraction field. This is in contrast to the case of argon \cite{li2016measurement} where it was found that the $QE$ in GAr is an order of magnitude higher than the $QE$ in LAr.

Future measurements will be performed to assess the stability and more precisely measure the $QE$ and work function of these gold photocathodes in LXe.

\section{\label{sec:level5}Conclusions}
A small drift cell for the study of electron drift properties in LXe was built and operated. A gold photocathode was back-illuminated and the released photoelectrons were investigated in a LXe environment and GXe. The longitudinal diffusion coefficient $D_L$ was measured in LXe as a function of drift field for st{the first time at fields below 1 kV/cm} a range of fields. The increase in $D_L$ with decreasing drift fields was qualitatively consistent with theoretical predictions \cite{robson1972thermodynamic,lowke1969theory,skullerud1969longitudinal}. Within the experimental uncertainty, no significant variation of $D_L$ with respect to temperature was observed. The field dependence of electron drift velocity in LXe and GXe agrees well with previously published values.

The use of calibrated charge bunches using a gold photocathode as a laser-driven electron source for \textit{in-situ} monitoring of electron lifetime is being further investigated for the nEXO design. It is important to assess the long term stability of the photocathode quantum efficiency as well as the precision and accuracy of the technique. A new cell will feature improved laser power monitoring, simultaneous  measurements from two photocathodes, and \textit{in-situ} source calibration crosschecks.

\section{Acknowledgment}
This work has been supported by the Offices of Nuclear and High Energy Physics within the DOE Office of Science, and NSF in the United States, by NSERC, CFI, FRQNT, and NRC in Canada, by IBS in Korea, by RFBR (18-02-00550) in Russia, and by CAS and NSFC in China. This work was supported in part by Laboratory Directed Research and Development (LDRD) programs at Brookhaven National Laboratory (BNL).

\section{References}
\bibliographystyle{elsarticle-num}
\bibliography{LongitudinalDiffusion_nEXOCollaboration}

\end{document}